\documentstyle[twoside,fleqn,npb,epsfig]{article}
\newcommand{\AmS}{{\protect\the\textfont2
  A\kern-.1667em\lower.5ex\hbox{M}\kern-.125emS}}

\newcommand{\beq}{\begin{equation}}
\newcommand{\eeq}{\end{equation}}
\newcommand{\bea}{\begin{eqnarray}}
\newcommand{\eea}{\end{eqnarray}}

\newcommand{\epm}{e^+e^-}

\newcommand{\ra}{\rightarrow}

\newcommand{\eemmbb}{e^+ e^- \ra \mu^+ \mu^- \bar{b} b}
\newcommand{\nn}{\nonumber}

\title{Towards precise predictions for the Higgsstrahlung
           at a linear collider\thanks{Work supported
           in part by the Polish State Committee for Scientific Research
           (KBN) under contract No. 2~P03B~045~23, by the European
           Community's Human Potential Program under contract
           HPRN-CT-2000-00149 Physics at Colliders and by DFG
           under Contract SFB/TR 9-03.}}

\author{F. Jegerlehner
   \address{Deutsches Elektronen-Synchrotron DESY, Platanenallee 6, 
            D-15738 Zeuthen, Germany}
        K. Ko\l odziej and T. Westwa\'nski
        \address{Institute of Physics, University of Silesia\\ 
          ul. Uniwersytecka 4, PL-40007 Katowice, Poland}}

\begin{document}

\begin{abstract}
We report on progress in our work on obtaining high precision standard model 
predictions for the Higgsstrahlung reaction at a linear collider.
\end{abstract}

\maketitle

\section{MOTIVATION}

Main production mechanisms of the standard model (SM) Higgs boson in 
$\epm$ collisions
at a linear collider \cite{NLC} are the Higgsstrahlung reaction
\bea
\label{eeZH}
                     e^+e^- \rightarrow  Z H 
\eea
and the $WW$ fusion process
\bea
\label{WW}
          e^+e^- \ra W^*W^* \ra  \nu_e\bar{\nu}_e H.
\eea

The cross section of (\ref{eeZH}) decreases as $1/s$ while
that of (\ref{WW}) grows as $\ln(s/m_H^2)$. Hence, while the Higgsstrahlung
dominates the Higgs boson production at low energies,
the $WW$ fusion process overtakes it at higher energies.
For small values of the Higgs boson mass, $m_H < 140$~GeV, as is the case 
considered in this talk, the Higgs boson dominantly decays into a 
$\bar{b} b$ quark pair. As the $Z$ boson of reaction (\ref{eeZH}) decays into 
a fermion--antifermion pair too, one actually observes the Higgsstrahlung 
through reactions with four fermions in the final state.

\begin{figure}[htb]
\vspace{15pt}
\includegraphics{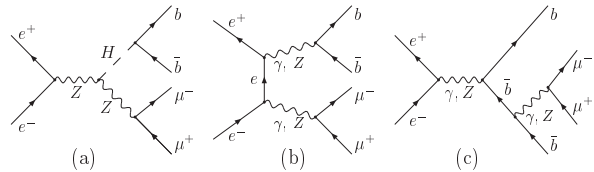}
\caption{Examples of the Feynman diagrams of reaction (\ref{bmmb}): 
(a) the double resonance `signal', (b) and (c) the 'background'
diagrams.}
\label{fig:diags}
\end{figure}

In the following we will concentrate on one specific four fermion channel
relevant for detection of (\ref{eeZH})
\bea
\label{bmmb}
 e^+(p_1)+e^-(p_2) \rightarrow  \mu^+(p_3)\!\!\!\! &+&\!\!\!\! \mu^-(p_4)\nn\\
                       \!\!\!\!&+&\!\!\!\! \bar{b}(p_5)+ b(p_6),
\eea
where the particle four momenta have been indicated in parentheses,
and the corresponding bremsstrahlung reaction
\bea
\label{bmmbg}
 e^+e^- \rightarrow \mu^+\mu^-\bar{b}b\gamma.
\eea
Reactions (\ref{bmmb}) and (\ref{bmmbg}) should leave a particularly clear 
signature in a detector.
To the lowest order of the SM, in the unitary gauge and neglecting the Higgs 
boson
coupling to the electron, reactions (\ref{bmmb}) and (\ref{bmmbg}) receive 
contributions from 34 and 236 Feynman diagrams, respectively. Typical 
examples of the Feynman diagrams of reaction (\ref{bmmb})  are 
depicted in Fig.~\ref{fig:diags}. The Higgsstrahlung `signal' diagram is 
shown in Fig.~1a,
while the diagrams in Figs.~1b and 1c represent typical `background' 
diagrams. The Feynman diagrams of reaction (\ref{bmmbg}) are obtained 
from those of reaction (\ref{bmmb}) by attaching an external photon line 
to each electrically charged particle line. 

In order to match the precision of data from a high luminosity 
linear collider, expected to be better than 1\%, it is necessary to include 
radiative corrections in the SM predictions. 
At the moment, however, it is not feasible to include the complete 
$\mathcal{O}(\alpha)$ corrections to a four--fermion reaction like
(\ref{bmmb}) into a Monte Carlo (MC) generator. Therefore it seems to 
be natural to apply the double 
pole approximation (DPA), as it had been done before in literature in case 
of the $W$-pair production  \cite{DPAWW} at LEP2.

To the lowest order of the SM, the cross sections of reactions (\ref{bmmb}) 
and (\ref{bmmbg}) can be computed with a program {\tt ee4fg} \cite{ee4fg}.
On the basis of {\tt ee4fg}, we have written a dedicated program 
{\tt eezh4f} that eventually should include all the leading radiative 
corrections
to the four--fermion reactions relevant for the Higgs boson production 
and decay through the Higgsstrahlung mechanism. However, 
up to now it includes the electroweak $\mathcal{O}(\alpha)$ radiative 
corrections to the on-shell reaction (\ref{eeZH}) of \cite{FJ1}
and the initial state radiation (ISR) QED corrections.
No corrections to $Z$ and Higgs decay widths, nor higher order ISR effects, 
nor any non-factorizable corrections are included in {\tt eezh4f} yet.

\section{OUTLINE OF THE CALCULATION}

The amplitude of (\ref{eeZH}) is represented in the following way
\bea
\label{MZH}
& & \hspace*{-1cm}
M_{ZH}\left(p_1,\sigma_1,p_2,\sigma_2,k_Z,\lambda_Z,k_H\right)=
\bar{v}\left(p_1,\sigma_1\right) \nn\\
&\times &\!\!\!\! {F}^{\mu}\left(p_1,p_2,k_Z,k_H\right)
\varepsilon_{\mu}^*\left(k_Z,\lambda_Z\right)u\left(p_2,\sigma_2\right),
\eea
where $p_1, p_2, k_Z, k_H$ and $\sigma_1, \sigma_2, \lambda_Z$ are the 
on-shell four momenta and helicities of the external particles.
Neglecting the electron mass, the amplitude 
${F}^{\mu}\left(p_1,p_2,k_Z,k_H\right)$
is decomposed in the following Lorentz covariant way 
\bea
\label{F}
\hspace*{-5mm}& & \hspace*{-6mm}{F}^{\mu}\left(p_1,p_2,k_Z,k_H\right)=
A_1\gamma^{\mu} + A_2\gamma^{\mu}\gamma_5 \\
\hspace*{-3mm}&+&\hspace*{-3mm}\left(A_{31}p_1^{\mu} 
                           + A_{32}p_2^{\mu}\right)/\!\!\!k_Z
+ \left(A_{41}p_1^{\mu} + A_{42}p_2^{\mu}\right)/\!\!\!k_Z\gamma_5,\nn
\eea
where we have changed a little the original notation with respect to 
that of Eq.~(2.5) of \cite{FJ1}. The six invariant amplitudes that do not
vanish in the limit $m_e \rightarrow 0$ are now denoted by
$A_i$ and $A_{ji}$, $i=1,2$, $j=3,4$.
In order to fix normalization we give here the two amplitudes, $A_i^{(0)}$, 
$i=1,2$, that are nonzero in the lowest order. They read
\bea
\label{ai0}
A_i^{(0)}=\frac{g_{HZZ}}{s-m_Z^2+i m_Z \Gamma_Z}\;{a_i},
\eea
with the vector and axial-vector coupling of the $Z$ to leptons
\bea
\label{Zll}
a_1=\frac{4s_W^2-1}{4s_Wc_W}\;e_W, \qquad a_2=\frac{e_W}{4s_Wc_W}
\eea
and the $Z$--Higgs boson coupling $g_{HZZ}=e_Wm_Z/(s_Wc_W)$;
the electric charge $e_W=\left(4\pi\alpha_W\right)^{1/2}$ 
and electroweak mixing angle $\theta_W$ are fixed by
\beq
\label{awsw}
\alpha_W=\sqrt{2} G_{\mu} m_W^2 s_W^2/\pi, \;\;
                s_W^2=1-m_W^2/m_Z^2,
\eeq
where $m_W$ and $m_Z$ are the physical masses of the $W$ and $Z$ boson.
This choice, which is exactly equivalent 
to the $G_{\mu}$-scheme of \cite{FJ1}, is in the program referred to
to as the {\em `fixed width scheme'} (FWS).
We have introduced a shorthand notation $s_W=\sin\theta_W$ and 
$c_W=\cos\theta_W$ in Eqs.~(\ref{Zll}--\ref{awsw}).

Amplitudes $A_i$, $i=1,2$, of Eq.~(\ref{F}) are split into the Born 
part $A_i^{(0)}$ of Eq.~(\ref{ai0}) and the one loop corrections in the
following way
\bea
\label{Ai}
A_i=A_i^{(0)}+\delta A_{i{\rm QED}} +\delta A_i, \qquad i=1,2,
\eea
where $\delta A_{i{\rm QED}}$ denotes the QED infra-red (IR) divergent
and $\delta A_i$ the IR finite part of the one-loop electroweak 
corrections.
The IR divergent part $\delta A_{i{\rm QED}}$ 
is factored out and combined with the soft bremsstrahlung
correction to (\ref{eeZH}) resulting in a QED initial state
radiation correction factor $C_{\rm QED}^{\rm ISR}$ \cite{FJ1}
that can be split into a universal cut-off dependent and non-universal 
constant part
\bea
\label{brem}
C_{\rm QED}^{\rm ISR}\left(s,E_{\rm cut}\right)=
C_{\rm QED}^{\rm univ.}\left(s,E_{\rm cut}\right) +
C_{\rm QED}^{\rm non-univ.},
\eea
with
\bea
\label{bremu}
C_{\rm QED}^{\rm univ.}\left(s,E_{\rm cut}\right) = 
   \frac{e^2}{2\pi^2} \hspace*{-6mm}& &\left[\left(\ln\frac{s}{m_e^2}-1\right)
         \ln\frac{2E_{\rm cut}}{\sqrt{s}}\right.\nn\\
        & & \hspace*{13mm}\left.+\frac{3}{4}\ln\frac{s}{m_e^2}\right],
\eea
and
\bea
\label{bremnu}
C_{\rm QED}^{\rm non-univ.} = 
   \frac{e^2}{2\pi^2} \left(\frac{\pi^2}{6}-1\right).
\eea
The electric charge $e=\left(4\pi\alpha_0\right)^{1/2}$ in 
Eqs.~(\ref{bremu}--\ref{bremnu}) is given in terms of $\alpha_0$ in 
the Thomson limit.

The one-loop amplitudes 
\bea
\label{ola}
\delta A_i=\delta A_i(s,\cos\theta),\quad
A_{ji} = A_{ji}(s,\cos\theta)
\eea
are complex functions of
$s=(p_1+p_2)^2$ and $\cos\theta$, $\theta$ being the Higgs boson production 
angle with respect to the initial positron beam in the centre of mass
system (CMS). 
They are computed numerically with {\tt eezh4f} making use of {\tt FF 2.0}, 
a package to evaluate one-loop 
integrals by G.~J. van Oldenborgh \cite{FF}. 

As the computation of the one-loop electroweak amplitudes of Eq.~(\ref{ola})
slows down the MC integration substantially, a simple
interpolation routine has been written that samples the amplitudes at
a few hundred values of $\cos\theta$
and then the amplitudes for all intermediate values of $\cos\theta$ are
obtained by a linear interpolation. This gives a tremendous gain in  
the speed of computation, while there is practically
no difference between the results obtained with the interpolation
routine and without it.

The one-loop virtual and soft bremsstrahlung corrected
matrix element squared of (\ref{bmmb}) in the double pole approximation
(DPA) reads
\bea
& & \hspace*{-1cm}
\left|M_{\rm virt + soft}^{\rm ISR}\right|^2 = \left|M^{(0)}\right|^2 
  \left(1+{C_{\rm QED}^{\rm univ.}}\right) \\
\hspace*{-3mm}&+& \hspace*{-3mm}\left|{M^{(0)}_{DPA}}\right|^2 
                {C_{\rm QED}^{\rm non-univ.}}
+ 2{\rm Re}\left({M^{(0)^*}_{DPA}\delta M_{DPA}}\right),\nn
\eea
where $M^{(0)}$ is the lowest order matrix element of (\ref{bmmb}) and
the QED correction factors are given by Eqs.~(\ref{bremu}) and (\ref{bremnu});
the lowest order matrix
element $M^{(0)}_{DPA}$ and the one-loop correction $\delta M_{DPA}$ 
in the double pole approximation
for the $Z$ and Higgs boson are given by
\bea
\label{m0}
M^{(0)}_{DPA} \!\!\!&=&\!\!\! \bar{v}\left(p_1\right)
\left(A_1^{(0)}\gamma_{\mu} + A_2^{(0)}\gamma_{\mu}\gamma_5\right)
           u\left(p_2\right)\nn\\
\!\!\!&\times& \!\!\! {P}_Z^{\mu} \; {P}_H\\
\label{dm}
\delta M_{DPA}\!\!\!&=&\!\!\!\bar{v}\left(p_1\right)\left[
\delta A_1\gamma_{\mu} + \delta A_2\gamma_{\mu}\gamma_5 \right.\nn\\
\!\!\!&+&\!\!\! \left(A_{31}p_{1\mu} + A_{32}p_{2\mu}\right)/\!\!\!k_Z \nn\\
\!\!\!&+&\!\!\! \left. \left(A_{41}p_{1\mu} + A_{42}p_{2\mu}\right)
            /\!\!\!k_Z\gamma_5\right] u\left(p_2\right)\nn\\
\!\!\!&\times&\!\!\! {P}_Z^{\mu} \; {P}_H.
\eea
The pole factors ${P}_Z^{\mu}$ and ${P}_H$ in 
Eqs.~(\ref{m0}) and (\ref{dm}) are given by
\bea
\label{pfs}
{P}_Z^{\mu} &=&
\frac{\bar{u}\left(k_4\right)\left(a_1\gamma^{\mu}
         +a_2\gamma^{\mu}\gamma_5\right) v\left(k_3\right)}
     {\left(p_3+p_4\right)^2-m_Z^2+im_Z\Gamma_Z},\nn\\
{P}_H &=& \frac{g_{Hbb} \bar{u}\left(k_6\right)v\left(k_5\right)}
                {\left(p_5+p_6\right)^2-m_H^2+im_H\Gamma_H},
\eea
with the $Z$--$\mu$ couplings $a_1,a_2$ given by Eq.~(\ref{Zll}) and 
the Higgs--bottom-quark coupling $g_{Hbb}=-e_W/(2s_W)(m_b/m_W)$.

The projected momenta $k_Z$ and $k_i$, of Eqs.~(\ref{dm}) and (\ref{pfs}) 
are obtained from the four momenta $p_i$, $i=3,...,6$, of the final state 
fermions of reaction (\ref{bmmb}) with 
the following projection procedure.

First the on-shell momentum and energy of the Higgs and $Z$ boson in the 
CMS are found
\bea
\left|\vec{k}_{H}\right|=
\frac{\lambda^{1/2}\left(s,m_Z^2,m_H^2\right)}{2s^{1/2}},\quad
\vec{k}_{H}=\left|\vec{k}_{H}\right|
\frac{\vec{p}_{5}+\vec{p}_{6}}
                 {\left|\vec{p}_{5}+\vec{p}_{6}\right|}, \nn
\eea
\bea
E_{H}=\sqrt{\vec{k}_{H}^2 + m_H^2}, \;\;\; \vec{k}_{Z}=-\vec{k}_{H}, \;\;
E_{Z}=\sqrt{s}-E_{H}.\nn
\eea

Denote four momenta $p_5$ and $p_3$ of reaction (\ref{bmmb})
that are boosted to the rest frame of the 
Higgs and $Z$, respectively, by $p'_5$ and $p'_3$. The projected four 
momenta $k'_i$, $i=5,6$, in the rest frame of the Higgs 
are obtained with
\bea
\left|\vec{k'}_{5}\right|\!\!\!&=&\!\!\!
\frac{\lambda^{1/2}\left(m_H^2,m_5^2,m_6^2\right)}{2m_H}, \quad
\vec{k'}_{5}=
\left|\vec{k'}_{5}\right|\frac{\vec{p'}_5}{\left|\vec{p'}_{5}\right|},\nn\\
\vec{k'}_{6}\!\!\!&=&\!\!\!-\vec{k'}_{5},\qquad
{E'}_{i}=\sqrt{\vec{k'}_{i}^2 + m_i^2}.\nn
\eea
Similarly one obtains $k'_3$ and $k'_4$ in the rest frame of the $Z$.

Four momenta $k'_i$, $i=3,...,6$, are then boosted back 
to the CMS giving the projected four momenta $k_i$, 
$i=3,...,6$ of $\mu^+, \mu^-, \bar{b}$ and $b$ that satisfy
the necessary on-shell relations
\bea
k_3^2=k_4^2\!\!\!&=&\!\!\!m_{\mu}^2, \qquad k_5^2=k_6^2=m_{b}^2,\nn\\
\left(k_3+k_4\right)^2\!\!\!&=&\!\!\!m_Z^2, 
\qquad \left(k_5+k_6\right)^2=m_H^2.\nn
\eea
The actual value of $\cos\theta$ in Eq.~(\ref{ola}) is given by 
$\cos\theta=k_H^3/|\vec{k}_H|$. 
The described projection procedure is not unique. As the Higgs boson width
is small,
the ambiguity between different 
possible projections is mainly related to the off-shellness of the $Z$ 
boson and is of the order of $\alpha\Gamma_Z/(\pi m_Z)$.

\section{NUMERICAL RESULTS}

In this section, we present some numerical results for (\ref{bmmb})
which will be one of the best detection channels of 
the Higgsstrahlung reaction (\ref{eeZH}) at the linear collider. 

The computation has been performed with a program {\tt eezh4f}
in the FWS with the following set of initial
SM electroweak physical parameters \cite{PDG}: 
\bea
\label{params1}
m_W\!\!\!&=&\!\!\!80.423\; {\rm GeV},  \quad \Gamma_W=2.118\; {\rm GeV},\\
m_Z\!\!\!&=&\!\!\!91.1876\; {\rm GeV},\quad \Gamma_Z=2.4952\; {\rm GeV},\\
G_{\mu}\!\!\!&=&\!\!\!1.16639 \times 10^{-5}\;{\rm GeV}^{-2}, \\
\alpha_0\!\!\!&=&\!\!\!1/137.03599976.
\eea
The charged lepton masses in MeV are
\bea
\label{params3}
m_e\!\!\!&=&\!\!\!0.510998902,\quad m_{\mu}=105.658357,\nn\\
m_{\tau}\!\!\!&=&\!\!\!1776.99
\eea
and the heavy quark masses in GeV are
\bea
m_c=1.5,\quad m_t=174.3,\quad m_b=4.4.
\eea
The light quark masses in MeV are assumed at
\beq
\label{params5}
m_u=62,\quad m_d=83,\quad m_s=215,
\eeq
which together with $\alpha_s=0.133$ reproduces
the hadronic contribution to the running of the fine structure constant.
The Higgs boson mass is assumed to be $m_H=115$~GeV and its
width is calculated to the lowest order of the SM.

\begin{figure}[htb]
\vspace*{6cm}
\includegraphics{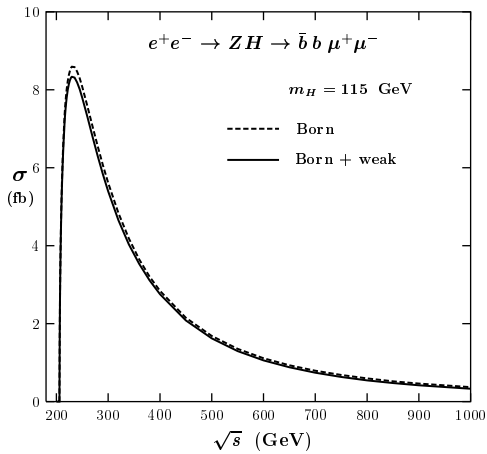}
\vspace*{-1cm}
\caption{The `signal' total cross section of reaction (\ref{bmmb}) 
         in the narrow width approximation as a function of
         CMS energy.}
\label{fig:ostot}
\end{figure}

\begin{figure}[htb]
\vspace*{6cm}
\includegraphics{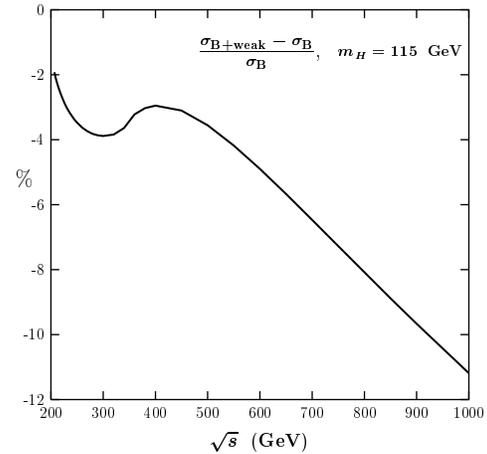}
\vspace*{-1cm}
\caption{The relative corrections corresponding to that of 
Fig.~\ref{fig:ostot}}
\label{fig:osrel}
\end{figure}

In Fig.~\ref{fig:ostot}, we plot the total 'signal' cross section of
(\ref{bmmb}) in the narrow width approximation, i.e. the cross section
of (\ref{eeZH}) multiplied by the corresponding branching ratios, for 
the Higgs and $Z$ boson, as a function of the CMS energy. The dashed curve 
shows the Born cross section while the solid curve shows the 
cross section including the one-loop electroweak corrections without
the QED initial state radiation correction factor $C_{\rm QED}^{\rm ISR}$ 
of Eq.~(\ref{brem}). The corresponding relative correction is plotted
in Fig.~\ref{fig:osrel}. The correction is of the order of a few per cent
for $\sqrt{s} < 500$~GeV and it grows almost linearly in the absolute
value for higher energies, most probably due to large Sudakov logs that 
enter through 
the wave function 
normalization factors. Once the corrections to the $Z$ and Higgs boson
decay are included, these large corrections combined with the wave
function renormalization of the $Z$ and Higgs decay width should cancel 
against the renormalization constant of the $Z$ and Higgs propagator.

\begin{figure}[htb]
\vspace*{6cm}
\includegraphics{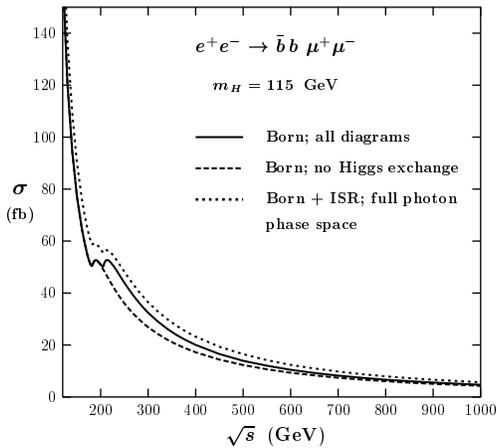}
\vspace*{-1cm}
\caption{Total cross section of (\ref{bmmb})
        as a function of CMS energy.}
\label{fig:tot}
\end{figure}

\begin{figure}[htb]
\vspace*{6cm}
\includegraphics{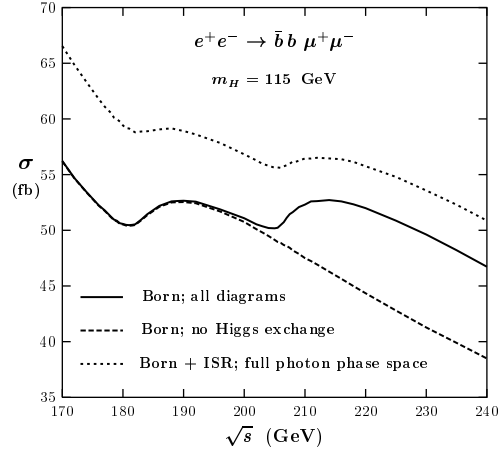}
\vspace*{-1cm}
\caption{Total cross section of (\ref{bmmb}) in the vicinity of 
         the $ZZ$ and $ZH$ resonances.}
\label{fig:tot1}
\end{figure}

In Fig.~\ref{fig:tot}, we compare results for the total cross sections 
of (\ref{bmmb}) 
as a function of the CMS energy, obtained with a complete set of the 
Feynman diagrams in the lowest order of the SM, including the Higgs boson 
(solid line) and without the Higgs boson exchange (dashed line). 
The $Z$--Higgs production signal is visible as the second bump
of the solid line, the first bump reflecting the double $Z$ production
resonance. The dotted line in Fig.~\ref{fig:tot} shows the 
$\mathcal{O}(\alpha)$ corrected cross section including the QED ISR
correction (\ref{bremu}) applied to the complete set of Feynman diagrams 
of (\ref{bmmb}), the initial state hard photon radiation integrated
over the full photon phase space and the electroweak radiative 
corrections to (\ref{eeZH}) in DPA. Fig.~\ref{fig:tot1} focuses
on the CMS energy region, where both resonances are present. We see
that the photon radiation smears the resonances, as expected.
The somewhat artificial large logarithmic correction of 
Fig.~\ref{fig:osrel} has practically no visible effect in 
Figs.~\ref{fig:tot} and \ref{fig:tot1} as the $\mathcal{O}(\alpha)$ is
dominated by the QED ISR correction.

\section{SUMMARY AND OUTLOOK}

The status of work on the high precision SM predictions for
the Higgsstrahlung reaction at the linear collider has been
reported.
The QED ISR corrections to $\eemmbb$ have been included.
The virtual one-loop weak corrections to (\ref{bmmb}) in the DPA have 
been included, too.
Work on inclusion of the corrections to $Z$ and Higgs 
decay widths and higher order corrections due to the QED ISR 
is in progress.
The non-factorizable corrections should be included, too.

\end{document}